\documentclass[aps,twocolumn,showpacs,superscriptaddress]{revtex4}
\usepackage{graphicx}
\begin{document}

\newcommand{\beq}{\begin{equation}}
\newcommand{\eeq}{\end{equation}}
\newcommand{\ben}{\begin{eqnarray}}
\newcommand{\een}{\end{eqnarray}}
\newcommand{\bea}{\begin{array}}
\newcommand{\eea}{\end{array}}
\newcommand{\om}{(\omega )}
\newcommand{\bef}{\begin{figure}}
\newcommand{\eef}{\end{figure}}
\newcommand{\leg}[1]{\caption{\protect\rm{\protect\footnotesize{#1}}}}
\newcommand{\ew}[1]{\langle{#1}\rangle}
\newcommand{\be}[1]{\mid\!{#1}\!\mid}
\newcommand{\no}{\nonumber}
\newcommand{\etal}{{\em et~al }}
\newcommand{\geff}{g_{\mbox{\it{\scriptsize{eff}}}}}
\newcommand{\da}[1]{{#1}^\dagger}
\newcommand{\cf}{{\it cf.\/}\ }
\newcommand{\ie}{{\it i.e.\/}\ }

\title{Irregular Dynamics in a One-Dimensional Bose System}

\author{G.P. Berman}
\affiliation{
Theoretical Division and CNLS, Los Alamos National
Laboratory, Los Alamos, New Mexico 87545}
\author{F.Borgonovi}
\affiliation{Dipartimento di
Matematica e Fisica, Universit\`a Cattolica, via Musei 41, 25121
Brescia, Italy}
\affiliation{I.N.F.M., Unit\'a di Brescia, Italy and
I.N.F.N., Sezione di Pavia, Italy}
\author{F.M. Izrailev}
\affiliation{
Instituto de F\'isica,
Universidad Aut\'onoma de Puebla, Apdo. Postal J-48, Puebla 72570,
M\'exico}
\author{A.Smerzi}
\affiliation{
Istituto Nazionale di Fisica per la Materia BEC-CRS\\
and Dipartimento di Fisica, Universit\`a di Trento, I-38050 Povo, Italy}

\begin{abstract}
We study many-body quantum dynamics of $\delta$-interacting bosons
confined in a one-dimensional ring. Main attention is payed to the
transition from the mean-field to Tonks-Girardeau regime using an
approach developed in the theory of interacting particles. We
analyze, both analytically and numerically, how the Shannon
entropy of the wavefunction and the momentum distribution depend
on time for a weak and strong interactions. We show that the
transition from regular (quasi-periodic) to irregular ("chaotic")
dynamics coincides with the onset of the Tonks-Girardeau regime.
In the latter regime the momentum distribution of the system
reveals a statistical relaxation to a steady state distribution.
The transition can be observed experimentally by studying the
interference fringes obtained after releasing the trap and letting
the boson system expand ballistically.
\end{abstract}

\date{today}
\maketitle

The study of the quantum many-body dynamics of a Bose-Einstein
condensate (BEC) is a
major challenge in the physics of trapped gases.
The mean-field Gross-Pitaevskii framework has
successfully described an impressive set of experimental data
\cite{dalfovo99}, but it fails when
genuine quantum many-body (QMB) correlations are important
(e.g., in quantum phase transitions
\cite{greiner02,fisher89},
in the design of quantum information devices \cite{rolston02} and
interferometers \cite{orzel01}, and near dynamically unstable
regions where quantum corrections appear on a
logarithmically short time-scale \cite{berman02}.

QMB effects also become particularly striking in low-dimensional systems,
where the competition between quantum fluctuations
and statistical properties is particularly enhanced.
Recent experimental achievements in effective one-dimensional (1D)
harmonically confined
quantum degenerate systems \cite{gorlitz01}
have stimulated several efforts
to understand their main properties \cite{blume02}.
The 1D regime can be achieved in optical/magnetic traps
when the radial degrees of freedom are frozen by tight
transverse confinement.
The Hamiltonian of the boson gas is given by the
Lieb-Liniger model \cite{lieb63}, where the two-body interaction is
assumed to be point--like. The thermodynamical properties of
this Hamiltonian, and its excitation spectra have been calculated
analytically by Lieb and Liniger \cite{lieb63}.
An unexpected feature, predicted by Girardeau \cite{girardeau60},
is the onset of fermionization when $n/g \to 0$
(here $n$ is the particle density,
and $g$ is the interatomic coupling constant which is inversely proportional
to the 1D interatomic scattering length \cite{olshanii98}).
In this Tonks-Girardeau (TG) regime, the
{\rm density} of the interacting bosons becomes identical
to that of non-interacting
fermions (while, of course, the wave-function keeps the
bosonic symmetry). On the other hand, in the opposite limit,
$n/g \to \infty$, the system is
described in the mean-field (MF)
approximation as a weakly interacting boson gas.
The crossover between these two regimes occurs near
$n / g \sim 1$. In this
region the MF approach breaks down and
more complicated two-body correlations become crucial.

While the thermodynamics of a 1D boson gas is fairly well
understood, systematic studies of many-body dynamical properties
of this system have been started only recently (see \cite{Lewen}
and references therein). In this context, one should mention
investigation of general quantum (beyond MF) properties of the
dynamics of trapped bosons, which can be considered as the next
frontier in BEC studies. In pursuit of this goal, in this Letter
we study numerically the quantum dynamics of $N$ interacting Bose
particles on a one-dimensional ring of length $L$. Our main result
is the discovery of a dynamical transition from regular to
(quantum) irregular dynamics when $n / g \to 1$. This transition
is accompained by a linear increase of the Shannon entropy of the
wave packets, that is directly related to the vanishing of the
interference fringes which occurs after releasing the confinement
and letting the bosons expand  ballistically.

Our model is specified by the Hamiltonian,
\begin{equation}
\displaystyle \hat{H} =
\sum_{k} \epsilon_k
\hat{n}_k +\frac{g}{2L} \sum_{k,q,p,r} \hat{a}_k^\dagger \hat{a}_q^\dagger
\hat{a}_p \hat{a}_r
\delta(k+q-p-r).
\label{ham}
\end{equation}
Here $\hat n_k=\hat{a}_k^{\dagger }\hat{a}_k$ is the occupation
number operator; $\hat a_{s}^{\dagger }$ and $\hat a_{s}$ are the
creation-annihilation operators; and $\epsilon_k = 4\pi^2
k^2/L^2$. The classical analog of this model, which is originated
from the nonlinear Shr\"odinger equation has been analyzed in
other papers. In the context of our study, the most interesting
results have been obtained in \cite{Lewen} where the interchange
between regular and irregular dynamics has been explored (without
the connection with the Tonks-Girardeau regime). Note, also, that
our model (\ref{ham}) is purely quantum and the Hamiltonian is
written in the operator form (see also \cite{pb98}). Specifically,
the occupation numbers for single-particle levels are not assumed
{\it apriori} to be very large. To analyze this model, below we
use the methods developed for quantum systems of interacting
particles with complex behavior (see, e.g. \cite{FI01b} and
references therein).

In what follows, we explore the situation when all bosons
initially occupy the single-particle level with the angular
momentum $k=0$ or, equivalently, at $t=0$ the system is in the
unperturbed ($g=0$) ground state $|\Psi_g\rangle$. Note that the
total momentum is conserved in time. Our main interest is the
evolution of the system for different values of the control
parameter $n / g$. This type of experiment can be realized since
the ratio $n / g$ can be manipulated by tuning the interatomic
scattering length.

It is convenient to label {\it single-particle states} $|k\rangle$
according to their momentum $k=0, \pm 1 , \pm 2... \pm m, ...$.
Then, any {\it many-body} state $|j\rangle$ can be represented as
$|...,n^j_{-m}, ... , n^j_0, ..., n^j_m,... \rangle$,
where $n^j_k$ represents the number of  particles in the single-particle
level characterized by the momentum $k$ of the $j$-th state.
In our numerical studies we need to take a finite number of particles
$N$ in a finite number $\ell=2m+1$ of single-particle states.
Clearly, the number $N$ of atoms and $\ell$ of levels
should be chosen in a consistent way
in order to have results which can be extrapolated to an arbitrary
number of atoms. In a 1D geometry
on a ring, $N$ particles define the smallest spacing, which corresponds
to the largest value of the momentum $m \approx N$.
In our numerical simulations, we satisfy the
latter relation when choosing changing values of $N$ and $m$.

The transition from the MF to the TG regime
can be understood using the following handwaving argument.
Having all particles initially in the lowest state
with $k=0$, let us estimate the strength of the interaction
necessary to move
two particles from the unperturbed ground state to the upper
(and lower) single particle levels $k=\pm m$. One
can expect that this interaction will result in an ergodic
filling (in time) of all single-particle states.
The energy required for the shift is approximately
$m^2/L^2 \approx  N^2/L^2$, and the matrix element of the interaction
between the corresponding states is  $<j | \hat{V} |j'> \sim g\sqrt{N(N-1)}/L$.
Equating these values one can obtain $n=N/L \sim g$ that is associated with
the crossover from the bosonic to fermionic regime.
Dynamically this crossover is reflected by a rapid depletion of
the fraction of particles in the lowest
single-particle state with $k=0$. Note that at $n / g \sim 1$, the ratio
$N_0 / N \sim {1 \over 2}$, with
$N_0 = \langle \hat{a_0}^\dagger \hat{a_0} \rangle$.

One of the most appropriate quantities to characterize the
dynamical properties of this system is the Shannon entropy of the
wave packet,
\begin{equation}
 S(t) =-\sum_j |\Psi_j (t) |^2 \ln |\Psi_j (t) |^2\,\,
\label{sh}
\end{equation}
Here $\Psi_j(t) = \langle j | \Psi(t)\rangle$ is the projection of
the wave function onto the noninteracting many-body basis. After
switching on the interaction, the wave function evolves according
to the total Hamiltonian, $\Psi(t) = e^ {-iHt} \Psi_g (0) $ and
spreads over the unperturbed basis.

\begin{figure}
\includegraphics[scale=0.37]{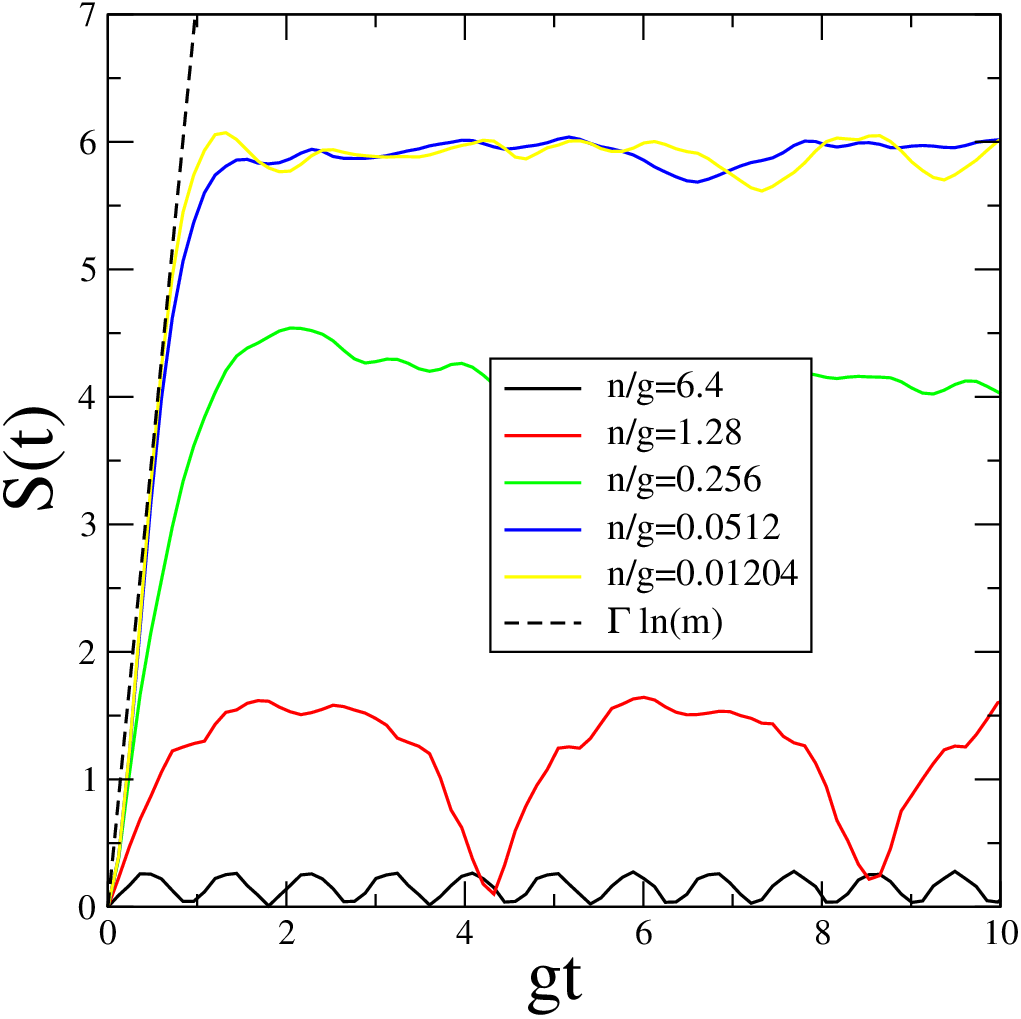}
\caption{
Entropy as a function of the rescaled time $gt$ for fixed $m=6$ and
different values of $n/g$. Data are given for $N=6$.
The saturation of $S(t)$ for small values of $n/g$ manifests the onset
of the Tonks-Girardeau regime. The dashed line corresponds to the
theoretical prediction (\ref{een}).
}
\label{becf3}
\end{figure}

The Shannon entropy, defined by Eq.(\ref{sh}) allows one to
estimate the effective number of unperturbed many-body states that
are involved in the dynamics due to the inter--particle
interaction, $N_{eff}(t) \approx \exp{\{S(t)\}}$. A few examples
of the time dependence of the entropy are reported in
Fig.\ref{becf3}. One can see that for $n/g \gg 1$ the entropy
oscillates in a regular way, while for $n/g \ll 1$ there is a
generic linear time dependence for a short time followed by a
saturation at later times.

While saturation is due to the finiteness of our Hilbert space,
the linear growth in time of the entropy was shown \cite{FI01b} to
be associated with the onset of chaos and thermalization in close
systems of randomly interacting particles (see details in
\cite{fi97} and {\cite{kota}). Specifically, for a strong enough
interaction between particles the linear increase of $S(t)$ is due
to an exponential increase in time of the number of many-body
basis states involved in the dynamics. Although our model is not
random, off-diagonal matrix elements determined by the particle
interaction  strongly fluctuate in time due to the different
number of particles occupying single particle levels. This fact
allows us to apply the results obtained in \cite{FI01b} for our
model.

For the case when the probability of remaining
in the originally excited state decreases exponentially, $W_0(t)
=\exp (-\Gamma t )$, it was found that the entropy increases linearly,
\begin{equation}
S(t) = \Gamma t \ln M \,\,,
\label{een}
\end{equation}
apart from a generic quadratic increase that occurs at small
times. Here the value of $\Gamma$ is determined by the
decay of the probability to stay in the unperturbed ground state,
and $M=m$ is the number of
many-body unperturbed states directly coupled to the initial state.
As was shown in \cite{FI01b}, the linear growth of entropy appears
in the regime for which $\Gamma$ is proportional to perturbation $g$.

We have carefully checked the predictions for the time-dependence
of $S(t)$ given in Ref.\cite{FI01b}.
In particular, we have confirmed
that the number of excited basis states grows in time
exponentially fast before saturation.
The saturation occurs because the total number of many-body states that we
had in our numerical simulation is finite. We also have found that
the probability $W_0(t)$ decays exponentially in time.
As a result, we were able to verify the increase of entropy given by
Eq.(\ref{een}) and indicated as a dashed line in Fig.\ref{becf3}.

\begin{figure}
\includegraphics[scale=0.37]{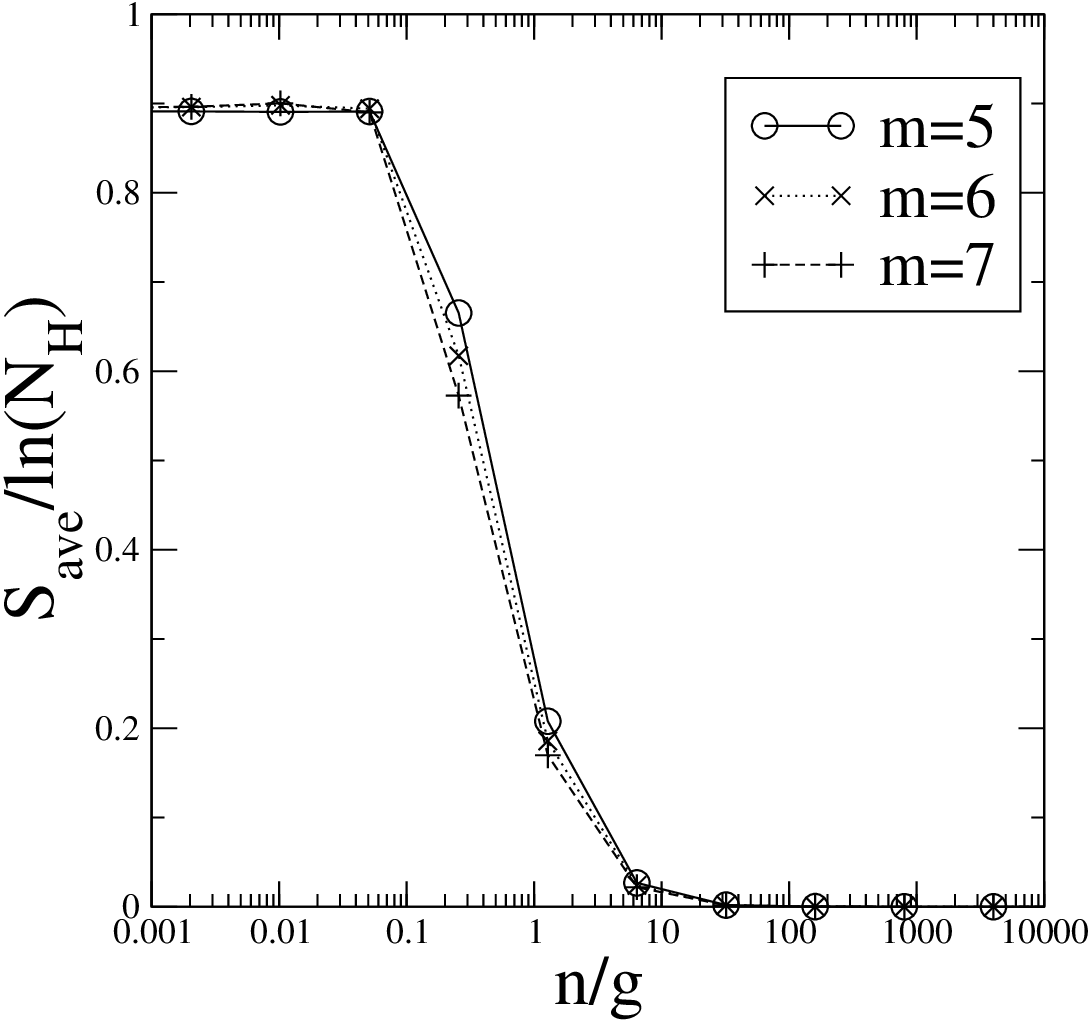}
\caption{
Rescaled entropy as a function of $n/g$ for $N=6$ particles
and different $m$ as indicated in the caption.}
\label{becf4}
\end{figure}

We summarize our main results in Fig.\ref{becf4}.
Here $S_{ave}$ is the mean value of the entropy averaged over time
after the saturation, and $N_H$ is the dimension of the Hilbert space (the
total number of many-body states defined by $m$ and $N$).
Entropy rescaling has been done in order to avoid its trivial
dependence on the size of the Hilbert space.
In this case the plateau for $n/g \ll 1$ is independent of
the choice of the number of particles and the number of levels.
One can see that the renormalized entropy is a quantity that is
very sensitive to the transition MF-TG regime.
Also, as one can see, slight variations in $m\approx N $ value does not
change the results. This indicates the universal character of
the entropy in the transition region. Note that the
limiting value of the renormalized
entropy is not  1, that would correspond to the prediction of
random matrix theory. This deviation clearly indicates the
influence of the dynamical nature of
the Hamiltonian.

We now discuss how our predictions can be tested experimentally.
Since the entropy of wave functions is not an easy quantity to measure
experimentally, we need to relate it to an
observable which can be measured directly.
In cylindrical coordinates the boson field operator (defined
on a ring of radius $R$
in the plane $z=0$) has the form,
$\hat{\psi} (\rho,\theta, z)  =  \hat{\delta}(z) \hat{\delta}
(\rho-R) \hat{\psi}(\theta)$.
Here the angle-dependent part can be expressed in terms
of plane waves,
$\hat{\psi} (\theta)  =  \sum_k \hat{a}_k e^{ik\rho \theta}$.
Fourier transforming,
$\hat{\chi} (\vec{p})  =  \int d_3 \vec{r} \
\hat{\psi} (\rho,\theta, z) \ e^{-i \vec{p} \cdot \vec{r}}$,
and taking into account that
$\vec{p} = p \hat{e}_y$
and  $\vec{p} \cdot \vec{r} = p \rho \sin \theta $, we have,
\begin{equation}
\hat{\chi} (p)  =   C \sum_k \hat{a}_k \int_\pi^\pi \ d\theta
e^{ikR\theta} e^{-ipR\sin(\theta)}
\label{al}
\end{equation}
where $C$ is a normalization constant.
Using  periodic boundary conditions,
$k= \frac{2\pi}{L} n = \frac{n}{R}$
with $n=0, \pm 1 \pm 2....$,
one obtains
\begin{equation}
\hat{\chi} (p) = C \sum_{n=-\infty}^{+\infty} J_n (pR) \ \hat{a}_n
\label{bf1}
\end{equation}
where $J_n(x)$ is the Bessel function of order $n$.

Expresssing the wave function at time $t$
in terms of the basis states
$|j\rangle \equiv |n^j_{-\infty},...n^j_0,... n^j_{+\infty} \rangle$,
we obtain an expression for the occupation number distribution
$n (p,t) =\langle \psi(t) | \hat{\chi}^\dagger (p) \hat{\chi} (p) |\psi(t)
\rangle$.
This quantity gives the probability density for finding
a boson with momentum $p$
along the $y$ axis at time $t$.
Using our expression (\ref{bf1}), we obtain,
\begin{equation}
n (p,t) = \sum_{j,j'} \bar{\psi}_j (t) \psi_{j'} (t) \sum_{l,s=-\infty}^{+\infty}
 J_l (pR) J_s (pR) \langle j| \hat{a}_l^\dagger
\hat{a}_s |j' \rangle
\label{de1}
\end{equation}

\begin{figure}
\includegraphics[scale=0.37]{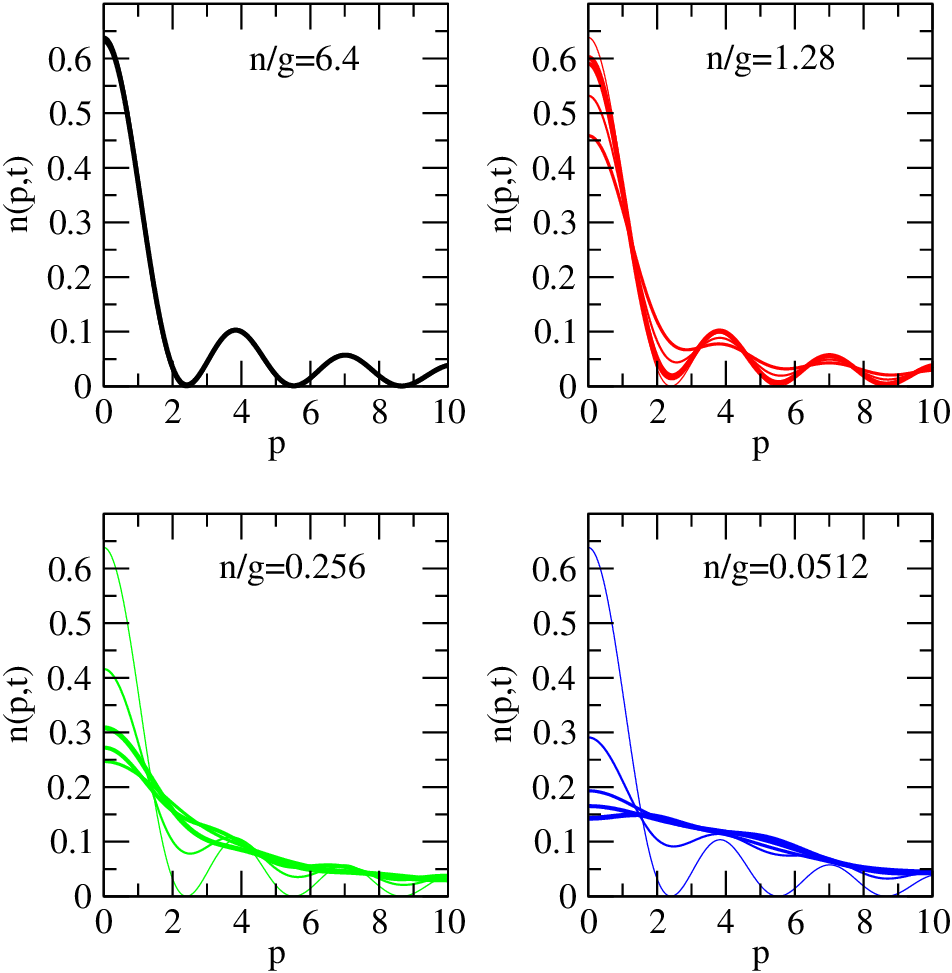}
\caption{
Particle density in the momentum representation for four
values of the interaction $n/g$,
and for different rescaled times $gt=0,0.2,0.4,0.6,0.8$
(lines are the thicker for later times).
}
\label{becf5}
\end{figure}

Since the total momentum is zero, the
selection rules,
$ \langle j| \hat{a}_l^\dagger
\hat{a}_s |j' \rangle  = n_l^j \delta_{j,j'} \delta_{l,s} $ hold,
where $n^j_l$ is the number of particles occupying
the single-particle level with momentum $l$
in the $j$-th basis state. Finally, we obtain,
\begin{equation}
n (p,t) = C
\sum_{j} | \psi_j (t) |^2 \sum_{l=-\infty}^{+\infty}
J_l^2 (pR) \ n_l^j
\label{de2}
\end{equation}
where $C^{-1} = N^{-1} \int \ dp \ n(p,t)$ .

\begin{figure}
\includegraphics[scale=0.37]{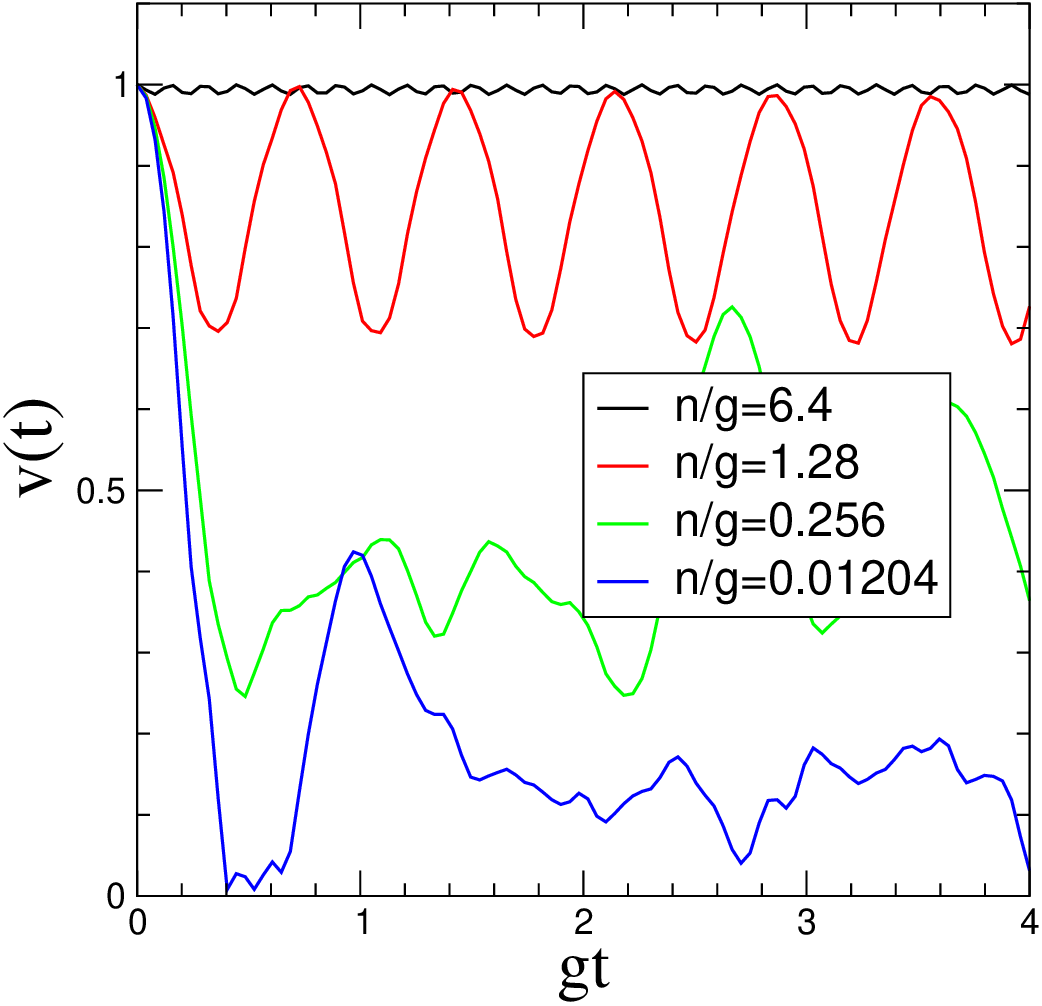}
\caption{
Visibility as a function of time, for four values of rescaled interaction
$n/g$. Here is $N=6$ and $m=6$.}
\label{becf6}
\end{figure}

Results are reported in Fig.\ref{becf5}. When $n/g >> 1$, the
momentum distribution does not practically change in time. All
particles remain mainly in the condensate during the evolution.
Close to the transition point $n/g \le 1$, the distribution begins
to be flat, without relaxing to an asymptotic distribution. In
this intermediate region, the occupation number distributions
reveal a mixture of very slow relaxation and strong oscillations
in time that was first observed \cite{pb98} in a similar model and
termed as ``non-ergodic behaviour of interacting bosons."

Contrary to this, for $n/g \ll 1$, one can see a clear relaxation
of the occupation number distribution to a steady state
distribution. In fact, the onset of relaxation to the steady
distribution means that the behavior of the model can be well
described by statistical methods \cite{FI01c}.

In order to characterize qualitatively the onset of relaxation in the TG regime,
we propose that one observe experimentally
the {\it fringe visibility}, defined as
\begin{equation}
v(t) = |I_{max}-I_{min}|/(I_{max}+I_{min})
\label{fri}
\end{equation}
Here we chose  $I_{max} = n(p=0,t)$,  $I_{min} = n(p_0,t)$
where $p_0 \simeq 2.41$ is the first zero of $J_0$. See Fig.\ref{becf6}.
This quantity allows one to distinguish between periodic
and aperiodic behavior of
the occupation number distribution. These different time behaviors
are shown in Fig.\ref{becf6}.

As one can see, the TG regime is
characterized by a fast decay of the visibility with subsequent
fluctuations that persist in time. These
fluctuations are additional evidence
of the statistical nature of the dynamics when the interparticle
interaction is strong.
When $n/g \gg  1$ (MF regime),
the visibility oscillates,
revealing collapses and revivals of the condensate fraction population,
as the entropy does in the same regime. See Fig.\ref{becf3}.
Therefore, there is
a strict correspondence between the dynamical evolution of the entropy and
the visibility of the momentum distribution.
The interfernce properties of a thermal Tonks gas localized on
one side of a ring
has been studied in \cite{das02}.

In conclusion, we have studied the dynamics of a Bose-Einstein condensate
model on a torus by paying attention to the role
of the interparticle interaction.
Specifically, we considered the situation in which initially
the ground state only, defined in the absence of
the interaction, is populated. By switching on the interaction we have found
different regimes depending on the strength of interaction.
The first regime (MF regime,
$n/g \gg1$) is characterized by regular dynamics
associated with periodic oscillations in time of the Shannon entropy
Eq.(\ref{sh}) and of the fringe visibility Eq.(\ref{fri}) of particle
density. In the Tonks-Girardeau regime, $n/g \ll 1$, the Shannon
entropy grows linearly in time in accordance with our analytical estimates.
Even if these estimates are obtained for completely
random models, they give remarkable agreement with our data.
Moreover, this approach allows one to understand the mechanism of the
transition from the mean field to Tonks gas regime.

We thank S. Giorgini, P. Pedri and L. Pitaevskii for valuable comments.
One of us (FMI) gratefully acknowledges the support by
CONACyT (Mexico) Grant No. 34668-E. This work has been partially
supported by the DOE, NSA, and ARDA.


\end{document}